\documentclass[conference,letterpaper]{IEEEtran}
\usepackage[T1]{fontenc}
\usepackage{mathtools}
\usepackage{cite}
\usepackage{graphicx}
\usepackage{booktabs}
\usepackage[checkfloat,nocheckfootnote,nospread]{flushend}
\usepackage[hidelinks]{hyperref}
\title{Stochastic Crew Placement for Winter Storm Restoration of Rural Distribution Feeders Considering Building Cooling Deadlines}
\author{\IEEEauthorblockN{Rajkumar Silwal, Shambhu Marasini and Bo Tu}
\IEEEauthorblockA{School of Electrical Engineering and Computer Science\\
University of North Dakota\\
Grand Forks, ND, USA\\
\{rajkumar.silwal, bo.tu\}@und.edu\\
shambhumarasini09@gmail.com}
}

\begin{document}
\maketitle
\begin{abstract}
Winter storm outages on rural distribution feeders may leave electrically heated houses cooling below acceptable indoor temperatures.
This paper proposes a two-stage stochastic mixed-integer linear program for pre-storm crew placement that considers the building cooling deadline.
Crews are placed before the storm, and repairs and reconnection are scheduled in each damage scenario to minimize the expected cost.
Specifically, outage house-hours beyond the deadline are penalized, and feeder voltage and loading constraints account for cold load pickup at each reconnection.
A case study on a synthetic rural feeder shows that the penalty shifts one of two crews toward a remote village, reducing expected house-hours beyond the deadline by 39 percent for 5.3 percent more outage house-hours.
With three crews, the penalty does not change crew placement.
Immediate reconnection would violate the voltage limit in four of six scenarios.
\end{abstract}


\section{Introduction}

\begin{figure}[!thb]
\centering
\includegraphics[width=1\columnwidth]{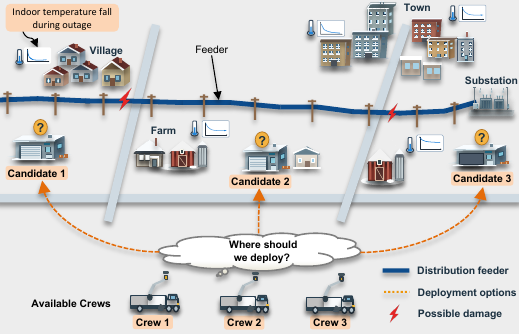}
\caption{Conceptual crew placement in a North Dakota community.
Distinct cooling curves illustrate thermal variation, while the model uses identical houses and a common deadline.}
\label{fig:concept}
\end{figure}

The feeder of interest in this paper is modeled on a rural North Dakota setting, a radial line from a substation in a town through farmland to a village and the farms beyond it.
Winter storms may damage overhead lines through ice accretion, high winds, and falling trees~\cite{keen2024}.
Without power, an electrically heated house cools at a rate set by its insulation and the outdoor temperature, so the indoor temperature can become unacceptable within several hours~\cite{ayyagari2020}.
However, the crews who repair the lines are few, distances are long, and roads are snow-covered after the storm, so a crew from the town may need hours to reach a remote village.
When a major storm is forecast, utilities therefore organize crews, trucks, and equipment before it arrives~\cite{eei2026}.
The decision is how many crews to prepare and where they should wait while the damage is still uncertain, as Fig.~\ref{fig:concept} illustrates.

Stochastic preparation models allocate resources before the damage is known and evaluate restoration over scenarios.
Pre-hurricane planning positions repair crews and resources to limit expected restoration cost and load shedding~\cite{arab2015}, a multi-commodity logistic model stages crews and equipment at distribution system depots~\cite{arif2020}, and distributed energy resources allocated before a storm are combined with post-storm dispatch and line repair scheduling~\cite{chang2018}.
Although these models place crews and resources before the damage is known, they weight an interruption by the load or customer-hours lost, independent of how long the customers can tolerate it.

After the damage is known, restoration models co-optimize repair crew routing with service restoration under uncertain repair times and demand~\cite{arif2018} and coordinate power and road repair crews when damaged roads delay access~\cite{zhu2022}.
Repair crew dispatch has also been coordinated with cold load pickup, the temporary increase in demand when customers regain power~\cite{pang2024}.
Although these models represent travel and repair in detail, cold load pickup enters only as a demand profile, no thermal deadline enters the schedule, and the crews are assumed to be in place.

Building thermal response has also entered resilience studies.
Thermal inertia of buildings and heating networks provides flexibility in the post-disaster preparation and recovery of a multi-energy distribution system~\cite{li2023}, substitution between electric and thermal demand adds flexibility to urban integrated energy system scheduling~\cite{zhu2026}, building response and road repair enter integrated electricity and transportation restoration~\cite{li2026}, and resilience assessment in extreme cold couples temperature, equipment failures, and heating demand~\cite{sheng2025}.
Although these works bring thermal response into resilience studies, they address post-event operation or assessment, and none of them connects it to pre-storm crew placement.

What matters most is whether each group of houses is reconnected before it becomes too cold, which depends on where the crews waited before the storm.
This paper addresses that gap by asking how the cooling deadline and an additional cost for the hours beyond it should change pre-storm crew placement.
The answer matters for the rural electric cooperatives of the northern Great Plains, where a December 2023 ice storm left members of one North Dakota cooperative without power for up to five days~\cite{npec2024}.

This paper proposes a two-stage stochastic mixed-integer linear program that links crew placement to a deadline from a building cooling model.
The first stage places crews at candidate waiting points, and the second schedules repairs and the reconnection of load blocks in each damage scenario.
The objective combines the preparation cost, the expected outage house-hours, and an additional cost for the house-hours beyond the cooling deadline.
The contributions are threefold.
First, a stochastic crew placement model with the cooling deadline and an additional cost for the hours beyond it.
Second, a continuous-time reconnection model that enforces feeder voltage and loading limits at every switch-on, with the warm-up demand of cold load pickup from earlier blocks included.
Third, a case study showing when the deadline changes the placement, how that depends on outdoor temperature and crew count, and why the switch-on check matters.

\section{Problem Formulation}\label{sec:formulation}

We formulate crew placement and restoration as a two-stage stochastic program.
The fixed radial feeder $G\coloneqq(\mathcal{N},\mathcal{E})$ has node set $\mathcal{N}$, line sections $\mathcal{E}$, and substation $0$ as its only source.
Each node $i\in\mathcal{N}\setminus\{0\}$ represents a load block of $N_i$ identical electrically heated houses that lose and regain power together.
Up to $M$ identical crews can wait at candidate waiting points $\mathcal{L}$, and preparing and sending one crew to point $\ell$ costs $C_\ell$.
All outages start at time zero, and times are in hours.

Scenario $s\in \mathcal{S}$ has damaged sections $\mathcal{F}_s\subseteq \mathcal{E}$ and probability $\pi_s$, with $\sum_{s\in \mathcal{S}}\pi_s=1$.
Only damage varies across scenarios.
Travel and repair durations are fixed, all roads and repair sites remain accessible, and outdoor temperature is constant during an outage.
At time zero, all damage is known, crews depart, and the restoration schedule is chosen.
Through Section~\ref{sec:limits}, we omit $s$ and write $\mathcal{F}$ for $\mathcal{F}_s$.

A damaged section disconnects the blocks behind its protective device until every damaged section assigned to that device is repaired.
For block $i$, $\mathcal{D}_i\subseteq \mathcal{F}$ lists the damaged sections that must be repaired to restore its supply.
The interrupted set is $\mathcal{I}\coloneqq\{i\in\mathcal{N}\setminus\{0\}\mid \mathcal{D}_i\neq\emptyset\}$.
Other blocks retain service.
Once their supply paths are available, blocks can be switched on independently through remotely controlled devices, without crew time or delay, and remain connected afterward.

\subsection{Building Cooling and Warm-up}\label{sec:cooling}
Every house starts at indoor temperature $\theta^0$ above the planning threshold $\underline\theta$ and has positive thermal time constant $\tau$.
The outdoor temperature $\theta_{\rm a}$ stays below $\underline\theta$.
With heating off and other heat gains neglected, the indoor temperature $\theta(h)$ after $h$ hours without power is~\cite{hao2015}
\begin{equation}
\theta(h)=\theta_{\rm a}+(\theta^0-\theta_{\rm a})\exp(-h/\tau).
\label{eq:cooling}
\end{equation}
It reaches the threshold at the cooling deadline
\begin{equation}
d\coloneqq\tau\ln\frac{\theta^0-\theta_{\rm a}}{\underline\theta-\theta_{\rm a}}.
\label{eq:deadline}
\end{equation}
After reconnection, each house draws its normal demand $p$ plus an additional heating demand $\Delta p$ for $\overline\delta$ hours, then $p$ alone.
This cold load pickup can overload lines or violate voltage limits when several blocks warm up at once.
The cooling equation alone does not determine the demand or duration of reheating, so for simplicity we define $\Delta p$ and $\overline\delta$ as fixed case inputs, used for every house regardless of outage length.
Indoor temperature after reconnection is not modeled.

\subsection{Crew Placement and Repair}\label{sec:model-constraints}
Before the storm, $x_\ell$ crews are placed at waiting point $\ell\in \mathcal{L}$, with
\begin{equation}
\sum_{\ell\in \mathcal{L}}x_\ell\leq M.
\label{eq:assignment}
\end{equation}
Crew routes start at waiting points and visit repair sites identified by their damaged-section labels.
For $a\in \mathcal{L}\cup \mathcal{F}$, $k\in \mathcal{F}$, and $a\neq k$, the binary variable $g_{ak}$ equals $1$ if the trip from $a$ to $k$ is selected and $0$ otherwise.
The routing constraints are
\begin{equation}
\begin{gathered}
\sum_{a\in(\mathcal{L}\cup \mathcal{F})\setminus\{e\}}g_{ae}=1,\qquad \forall e\in \mathcal{F},\\
\sum_{k\in \mathcal{F}\setminus\{e\}}g_{ek}\leq1,\qquad \forall e\in \mathcal{F},\\
\sum_{k\in \mathcal{F}}g_{\ell k}\leq x_\ell,\qquad \forall\ell\in \mathcal{L}.
\end{gathered}
\label{eq:routing}
\end{equation}
They assign each repair once, allow at most one next repair, and limit departures to the crews placed at each point.

Travel takes $\delta_{\ell k}$ from waiting point $\ell$ to repair site $k$ and $\delta_{ek}$ from repair site $e$ to repair site $k$.
Repair $k$ occupies one crew without interruption for positive duration $\delta_k$ and finishes at $t_k$.
For departures at time zero and no crew idle time, selected trips require, for $k\in \mathcal{F}$, $\ell\in \mathcal{L}$, and $e\in \mathcal{F}\setminus\{k\}$,
\begin{equation}
\begin{aligned}
g_{\ell k}=1&\ \Longrightarrow\ t_k=\delta_{\ell k}+\delta_k,\\
g_{ek}=1&\ \Longrightarrow\ t_k=t_e+\delta_{ek}+\delta_k.
\end{aligned}
\label{eq:timing}
\end{equation}
Positive repair durations exclude cycles in~\eqref{eq:timing}, so with~\eqref{eq:routing} every route starts at a staffed waiting point.
Idle time is unnecessary because earlier repair completion never forces earlier reconnection.

\subsection{Reconnection}\label{sec:restoration}
The outage cost depends on when each block regains power, and upstream supply or feeder limits can delay reconnection after all required repairs are complete.
We therefore introduce block reconnection times to link the repair schedule in~\eqref{eq:timing} to outage duration and delays beyond the cooling deadline.
For interrupted block $i\in \mathcal{I}$, let $T_i$ be its reconnection (switch-on) time and $u(i)$ its immediately upstream node.
Set $T_0=0$ and $T_i=0$ for blocks retaining service.
For every $i\in \mathcal{I}$, reconnection requires upstream power and completion of all repairs in $\mathcal{D}_i$,
\begin{equation}
T_i\geq T_{u(i)},\qquad T_i\geq t_e\quad\forall e\in \mathcal{D}_i.
\label{eq:restoration}
\end{equation}
Feeder limits may delay $T_i$ beyond these bounds while crews continue repairing independently.
The outage cost counts all time until $T_i$.
The late outage duration $q_i\coloneqq\max\{0,\,T_i-d\}$ counts time without power beyond $d$ and excludes warm-up after reconnection.

\subsection{Feeder Limits at Switch-on}\label{sec:limits}
Blocks still warming up draw the additional demand, so at each switch-on we add up the demand of every block with power and check it against the feeder limits.
At each switch-on $T_i$, $i\in \mathcal{I}$, binary flags $o_{ij}$ and $f_{ij}$ count block $j$ as connected and its warm-up as complete, respectively.
For distinct $i,j\in \mathcal{I}$,
\begin{equation}
\begin{aligned}
o_{ij}=0&\ \Longrightarrow\ T_j-T_i\geq\varepsilon,\\
f_{ij}=1&\ \Longrightarrow\ T_i-T_j\geq\overline\delta.
\end{aligned}
\label{eq:order}
\end{equation}
Here $\varepsilon$ is a small positive timing margin no longer than $\overline\delta$.
For the block being reconnected, $o_{ii}=1$ and $f_{ii}=0$, and both flags are $1$ for blocks retaining service.
These rules imply $f_{ij}\leq o_{ij}$ and allow demand overcounting but prevent undercounting.

Neglecting losses, radial line flow equals total downstream demand.
For section $e\in \mathcal{E}$, let $\mathcal{B}_e$ contain its downstream blocks, including the block at its receiving node.
Blocks counted as connected contribute normal demand, and those counted as still warming up add the additional demand.
The flow on section $e$ used in the check at $T_i$, $i\in \mathcal{I}$, is
\begin{equation*}
P_{ei}\coloneqq\sum_{j\in \mathcal{B}_e}N_j\,\big[p\,o_{ij}+\Delta p\,(o_{ij}-f_{ij})\big].
\end{equation*}
These flows determine voltage through the lossless linear radial power-flow approximation, the linearized DistFlow model~\cite{baran1989,arif2018}.
For a common lagging or unity power factor, $z_k$ is section $k$'s per-unit resistance plus its per-unit reactance times the tangent of the power-factor angle.
Thus $2z_kP_{ki}$ estimates that section's squared-voltage drop.
With source voltage $1$ per unit and path sections $\mathcal{U}_j$ from the substation to node $j$, the estimated squared voltage at node $j$ is $1-2\sum_{k\in \mathcal{U}_j}z_kP_{ki}$, and this estimate keeps the model linear.

Let $\overline P_e$ be section $e$'s active-power limit and $\underline v$ the minimum allowed squared voltage.
For every $e\in \mathcal{E}$, $j\in\mathcal{N}\setminus\{0\}$, and $i\in \mathcal{I}$, require
\begin{equation}
P_{ei}\leq\overline P_e,\qquad
2\sum_{k\in \mathcal{U}_j}z_kP_{ki}\leq1-\underline v.
\label{eq:limits}
\end{equation}
Between switch-ons, demand only falls as warm-ups end, so no further checks are needed.
We assume that reconnecting blocks one at a time from the substation outward, after the preceding block has warmed up, passes all checks.

\subsection{Two-Stage Stochastic Program}\label{sec:twostage}
Crew counts $x_\ell$ are chosen before damage is known and shared across scenarios.
After observing $s$, the second stage chooses the restoration schedule, including repair routes, repair completion times, and block reconnection times.
These variables carry superscripts $s$.
Scenario sets are $\mathcal{F}_s$, $\mathcal{I}_s$, and $\mathcal{D}_{i,s}$.
Each outage house-hour costs the price $w_{\rm o}\geq0$, plus the late price $w_{\rm l}\geq0$ for each house-hour beyond $d$.
Preparation is charged once and scenario costs are probability-weighted,
\begin{equation*}
\begin{aligned}
\min\quad&\sum_{\ell\in \mathcal{L}}C_\ell x_\ell+\sum_{s\in \mathcal{S}}\pi_s\sum_{i\in \mathcal{I}_s}N_i\big(w_{\rm o}T_i^s+w_{\rm l}q_i^s\big)\\
\text{s.t.}\quad&\eqref{eq:assignment},\\
&\eqref{eq:routing}\text{--}\eqref{eq:limits}\quad\forall s\in \mathcal{S},\\
&q_i^s\geq0,\quad q_i^s\geq T_i^s-d\quad\forall s\in \mathcal{S},\ i\in \mathcal{I}_s.
\end{aligned}
\end{equation*}
Crew counts are nonnegative integers, $g_{ak}^s$, $o_{ij}^s$, and $f_{ij}^s$ are binary, and $t_e^s,T_i^s$ are nonnegative.
A positive objective coefficient makes $q_i^s$ equal its larger lower bound.
Reported late house-hours always use $N_i\max\{0,T_i^s-d\}$, including at zero late price.

All scenarios can be combined in one model with common crew counts and separate restoration variables.
Because the scenarios share only the crew counts, the problem can also be solved by enumerating all placements, solving each placement's scenario problems independently, and choosing the lowest total of preparation cost and probability-weighted scenario costs.
The conditional rules in~\eqref{eq:timing} and~\eqref{eq:order} are written as bounded linear inequalities, with time bounds from travel, repair, and warm-up durations.

\section{Case Study}\label{sec:case}
\begin{figure}[!thb]
\centering
\includegraphics[width=\columnwidth]{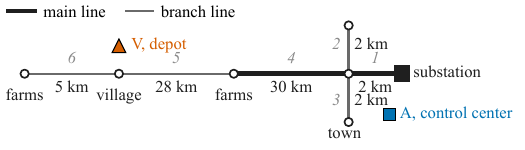}
\caption{Feeder schematic, not to scale, with section numbers in italics and lengths in km.}
\label{fig:feeder}
\end{figure}

We use a synthetic 24.9 kV rural feeder serving 680 houses in a town, a village 60 km west of the substation, and farms along the line (Fig.~\ref{fig:feeder}).
Line section $i$ terminates at node $i$, representing load block $i$.
Table~\ref{tab:inputs} lists houses per block and assumed active-power limits and repair times per line section.
Main lines (sections 1 and 4) and branch lines have resistances of 0.2 and 0.7 $\Omega$/km, respectively, and per-unit values use a 1~MVA base.

\begin{table}[!thb]
\caption{Load-Block House Counts and Line-Section Power Limits and Repair Times}
\label{tab:inputs}
\centering
\footnotesize
\setlength{\tabcolsep}{6pt}
\begin{tabular}{lrrr}
\toprule
Line section $e$ & Houses & $\overline P_e$ (MW) & $\delta_e$ (h) \\
\midrule
1, town main line & 160 & 8.0 & 2.0 \\
2, north branch line & 190 & 3.0 & 1.5 \\
3, south branch line & 190 & 3.0 & 1.5 \\
4, west main line & 20 & 2.0 & 4.0 \\
5, village line & 100 & 1.5 & 5.0 \\
6, beyond the village & 20 & 1.5 & 3.0 \\
\bottomrule
\end{tabular}
\end{table}

For the resistance-heated houses, we assume a normal demand of $p=5$ kW, or 0.005 per unit on the 1 MVA base, and an additional demand of $\Delta p=5$ kW for $\overline\delta=5.5$ h after reconnection, giving a total warm-up demand of 10 kW per house.
If the normal demand holds $\theta^0$ at $-20\,^\circ$C, the first-order model~\eqref{eq:cooling} with the heating on needs 4.8 h to return a house from $\underline\theta$ to $\theta^0$ with this demand, so $\overline\delta$ is not shorter than the warm-up of a house reconnected at the deadline.
A longer warm-up, which colder weather would bring, would lengthen the switch-on delays and add late house-hours, so the case understates the need for the check.
The power factor is 1, so $z_k$ in~\eqref{eq:limits} is the resistance of line section $k$.
With $\theta^0=21\,^\circ$C, $\underline\theta=10\,^\circ$C, and $\tau=20$ h, \eqref{eq:deadline} gives $d=6.2$ h at the outdoor temperature of $-20\,^\circ$C.
Fig.~\ref{fig:sweep} varies the outdoor temperature so that the deadline runs from 4.9 to 8.8 h, on both sides of the village's reconnection times from the two waiting points.
The source voltage is nominal, no voltage regulator is modeled, and the minimum allowed voltage magnitude is 0.95 per unit.

Two crews ($M=2$) can wait at the town control center (A) or the village depot (V), with equal preparation costs at both points, and both are always placed, so~\eqref{eq:assignment} holds with equality.
Each section has its own protective device and is repaired at its midpoint.
A drive covers the east-west plus north-south distance to that midpoint at 30 km/h on snow-covered roads, plus 0.3 h for loading and finding the fault.
That puts the village line 1.8 h from A and 0.8 h from V.
The six damage scenarios, listed by damaged section numbers followed by their probabilities, are $\{5\}$ 0.3, $\{5, 6\}$ 0.1, $\{4, 5\}$ 0.1, $\{4\}$ 0.1, $\{2, 3\}$ 0.3, and $\{3\}$ 0.1, chosen so that both town and rural damage matter.
Each outage house-hour costs $w_{\rm o}=\$2$, with an additional $w_{\rm l}=\$4$ for each house-hour beyond the cooling deadline, and $\varepsilon$ is one minute.
We solved each scenario problem with the open-source solver CBC, and the combined model, with 175 variables and 402 constraints, returns the same placement in under a second.

\section{Results}\label{sec:results}
Without the late price, the best placement keeps both crews at the town control center (A+A).
With it, the best placement puts one crew at the village depot (A+V).
Table~\ref{tab:results} compares these placements with the late price included in each restoration schedule.
Moving one crew to the village cuts expected late house-hours by 39 percent for 5.3 percent more outage house-hours.
In every scenario with the village line damaged, this move advances the village's reconnection from 6.8 to 5.8 h, bringing it before the 6.2 h deadline.
The tradeoff comes when both town branch lines are damaged.
The crew from V then reaches its branch line 1.8 h later than a second crew from A would, so 190 houses wait longer.

\begin{table}[!htb]
\caption{Expected House-Hours for Two-Crew Placements With the Late Price (Rounded to Whole Numbers)}
\label{tab:results}
\centering
\footnotesize
\setlength{\tabcolsep}{8pt}
\begin{tabular}{llrr}
\toprule
Switch-on check & Placement & Outage & Late \\
\midrule
With & A+A & 798 & 96 \\
With & A+V & 840 & 59 \\
\midrule
Without & A+A & 732 & 33 \\
Without & A+V & 774 & 0 \\
\bottomrule
\end{tabular}
\end{table}

\begin{figure}[!b]
\centering
\includegraphics[width=0.9\columnwidth]{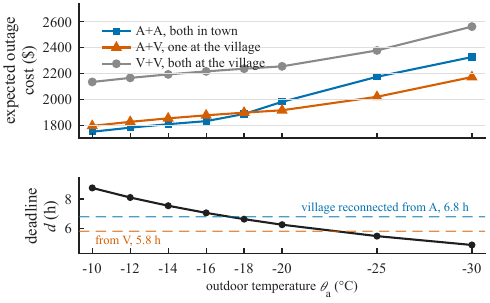}
\caption{Top, the expected outage cost of each placement with the late price, preparation cost excluded.
Bottom, the cooling deadline and the village's reconnection times from A and V.}
\label{fig:sweep}
\end{figure}

Fig.~\ref{fig:sweep} repeats the comparison at outdoor temperatures from $-10$ to $-30\,^\circ$C, with the deadline recomputed from~\eqref{eq:deadline} and $p$, $\Delta p$, and $\overline\delta$ held fixed.
At $-18\,^\circ$C and milder, A+A has the lowest expected cost.
At $-20\,^\circ$C and colder, A+V does.
Moving one crew to the village costs the same 42.6 expected outage house-hours at every temperature, \$85.2 at \$2 each.
It saves late house-hours worth \$74.6 at $-18\,^\circ$C, where the deadline is 6.6 h and the village's reconnection from A at 6.8 h is already late, and \$150.5 at $-20\,^\circ$C.
The saving passes the cost between those two temperatures.
Equivalently, A+V costs less than A+A once the late price exceeds about 1.1 times the outage price at $-20\,^\circ$C and 2.3 times at $-18\,^\circ$C.
V+V costs the most at every temperature, because the town's branch lines are then 60 km from the nearest crew.
With a third crew, two crews in town and one at the village costs least with or without the late price, so the late price changes the placement only when crews are scarce.

The comparison without the switch-on check shows why the feeder limits in Section~\ref{sec:limits} matter.
The same crew routes were kept, but every interrupted block was switched on at the earliest time~\eqref{eq:restoration} allows.
An alternating-current power flow, a backward/forward sweep~\cite{shirmohammadi1988}, checks voltages at every switch-on and end of warm-up, using reactances of 0.4 $\Omega$/km on the main lines and 0.5 $\Omega$/km on the branch lines.
Under either placement, early switch-on in the four scenarios with damage west of the town, on sections 4 to 6, lowers the minimum voltage to 0.940--0.941 per unit for the 5.5 h warm-up, below the 0.95 floor.
With the check, the minimum is 0.951--0.952 per unit.
The early states also fail the linear voltage check, while no line section exceeds its active-power limit.

\begin{figure}[!tbh]
\centering
\includegraphics[width=0.8\columnwidth]{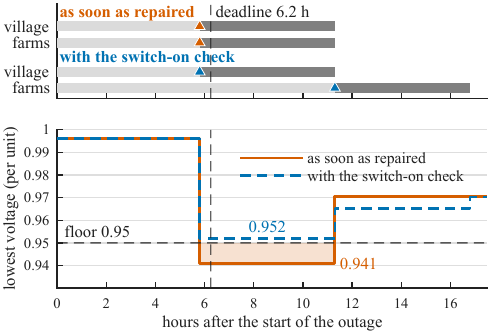}
\caption{Village-line damage under A+V.
Top, the village and the farms beyond it without power (light bars) and warming up (dark bars), as soon as repaired and with the switch-on check.
Bottom, the lowest voltage among the houses with power.
Triangles mark switch-on.}
\label{fig:voltage}
\end{figure}

Fig.~\ref{fig:voltage} shows the village-line fault under A+V.
The line is repaired at 5.8 h, but switching on the village and the farms beyond it together would lower the end voltage to 0.941 per unit.
With the check, the village comes on at 5.8 h and the farms wait until its warm-up ends at 11.3 h.
This delay leaves the farms late under either placement, accounting for most of the remaining late house-hours.
Across all scenarios, the delay adds 66 expected outage house-hours, the gap between the rows with and without the check in Table~\ref{tab:results}.

\section{Conclusion}\label{sec:conclusion}
This paper proposed a two-stage stochastic mixed-integer linear program that places repair crews before a winter storm, with the building cooling deadline priced in the objective and the feeder voltage and loading limits checked at every switch-on during cold load pickup.
In the case study, the deadline changes the placement when it falls before the reconnection time a large remote block gets from the control center, when the late house-hours a nearer crew saves are worth more than the outage house-hours the move costs elsewhere, and when crews are scarce.
Because outdoor temperature sets the deadline, the least-cost placement can change with the forecast temperature for the same feeder and damage scenarios.
The case study also shows that switching blocks on as soon as their repairs are complete can hold the voltage below its floor for the whole warm-up, so the switch-on check belongs in the restoration schedule itself.
These conclusions rest on identical houses, fixed travel and repair durations, a fixed warm-up demand, and six damage scenarios.

Future work will apply the model to utility feeder data with damage scenarios drawn from storm forecasts and line fragility models, add crew duty rules, manual switching, and uncertain repair times, and bring alternating-current limits inside the optimization.
A thermal model of reheating would also let the warm-up demand and the cost of late reconnection come from the same building model.

\section*{Acknowledgment}
The authors used both Claude (Anthropic) and ChatGPT (\mbox{OpenAI}) to help refine the illustrative figure (Fig.~2) and the language of this paper.
The authors reviewed all content and take full responsibility for it.

\bibliographystyle{IEEEtran}
\bibliography{references_v4}
\end{document}